\documentclass[12pt,floats]{iopart}
\usepackage{epsfig}

\def\no{\nonumber}
\def\a{\alpha}
\def\b{\beta}
\def\e{\epsilon}
\def\g{\gamma}

\def\t{\theta}

\def\be{\begin{equation}}
\def\bea{\begin{eqnarray}}
\def\eea{\end{eqnarray}}
\def\ee{\end{equation}}
\def\bi{\begin{itemize}}
\def\ei{\end{itemize}}

\def\d{\delta}

\def\vt{\vartheta}

\def\l{\left}
\def\r{\right}
\def\bn{\begin{enumerate}}
\def\en{\end{enumerate}}

\begin{document}

\title[Computational cost for detecting inspiraling binaries using a 
network...]{Computational cost for detecting inspiraling binaries using a 
network of laser interferometric detectors}
\author{Archana Pai \dag, Sukanta Bose \ddag \S, and Sanjeev Dhurandhar \dag}
\address{\dag Inter-University Centre for Astronomy and Astrophysics, Pune, 
India}
\address{\ddag Department of Physics and Program in Astronomy, Washington State
University, Pullman, WA 99164-2814, USA}
\address{\S Max Planck Institut f\"{u}r Gravitationphysik, Albert Einstein 
Institut, Am M\"{u}hlenberg 1, Golm, D-14476, Germany}

\ead{apai@iucaa.ernet.in, sukanta@wsu.edu, sanjeev@iucaa.ernet.in}

\begin{abstract}
We extend a coherent network data-analysis strategy developed earlier for 
detecting Newtonian
% the network of detectors for Newtonian chirp 
waveforms to the case of post-Newtonian (PN) waveforms. 
Since the PN waveform depends on the individual masses of the inspiraling
binary, the parameter-space dimension increases by
1 from that of the Newtonian case. We obtain the number of templates 
and estimate the computational costs for PN waveforms: For a lower mass limit 
of $1 M_{\odot}$, 
%the number of PN templates 
for LIGO-I noise, and with $3 \%$
maximum mismatch, the online computational speed requirement for 
single detector is a few Gflops; for a two-detector network it is 
hundreds of Gflops and for a three-detector network it is tens of Tflops. 
Apart from idealistic networks, we obtain results for realistic networks 
comprising of LIGO and VIRGO. Finally, we compare costs incurred in a 
coincidence detection strategy 
with those incurred in the coherent strategy detailed above.
\end{abstract}

%\submitto{\CQG}
\maketitle

\section{Introduction}
Close compact binaries are among the prime sources of gravitational waves that
hold promise for detection with upcoming laser interferometric detectors such 
as LIGO, VIRGO, GEO-600, TAMA, and AIGO. The back-reaction of radiated 
gravitational waves  results in an inspiral with an eventual merger of the two 
companions 
of the binary system. This adiabatic inspiral waveform has been accurately 
modeled upto 2.5 post-Newtonian order (PN) \cite{DIB01}. 
In an earlier work \cite{PDB01},
we developed a formalism for detecting inspiral waveform
with a network of detectors. The proposed analysis is of coherent nature
where the network is treated as a single detector and the data is combined
using the phase information optimally. In \cite{PDB01}, we used the
maximum likelihood detection (MLD) technique, which involves correlating the 
output of a network of detectors with the family of expected waveforms 
(or templates) and selecting the
maximum of the network likelihood ratio for decision making \cite{Hels}. 
To reduce computational costs
involved in searching over the space of source parameters, we succeeded in
analytically  maximizing
over 4 of these parameters, namely, the overall amplitude, initial
phase and the orientation angles of the binary orbit. The maximization
over the time of arrival (or, alternatively, over the time of final 
coalescence) of signal was carried out via FFTs. Estimates of computational
costs involved in searching over the source-direction 
angles and the chirp mass were obtained for the simplistic case of Newtonian 
waveforms. In this work, we extend the coherent network analysis to the 
more realistic case of PN waveforms. A restricted PN waveform depends on 
individual masses of the companions instead of the combined chirp mass.
This increases the number of parameters by one. We estimate, in general, the 
costs involved in searching over the masses as well as the source-direction
angles for realistic network configurations. Finally, we describe a 
coincidence network detection strategy and compare costs 
incurred in it with those in the coherent detection strategy.

\section{Restricted post-Newtonian signal at the network}
The signal $s^I(t)$ at the constituent $I$-th detector of the network 
is given by \cite{PDB01}
\be \label{primsig}
s^I(t) = 2 \kappa \>\Re \left[ ( E_{I}^{*} S^{I})e^{i\delta_c} \right]\ \ ,
\ee
where $\kappa$ is the overall amplitude that depends on 
the fiducial frequency $f_s$ and the
masses of the binaries. $\delta_c$ is the phase of the waveform at the time of
final coalescence. The extended beam-pattern functions of the $I$-th detector, 
$E_{I}$, depend on the source-direction angles, $\{\theta,\phi\}$, the orbit 
orientation angles, $\{\epsilon,\psi\}$, the $I$-th detector orientation, 
$(\a_{(I)},\b_{(I)},\g_{(I)})$, and the sensitivity, $g_{(I)}$, of
the detector to the incoming signal. Finally, $S^{I}(t)$ is a normalized
complex signal such that in the
stationary-phase approximation (SPA) its Fourier transform (FT) is
\be
{\tilde S}^I(f;t_c,\xi) 
= {2 \over {g_{(I)}}} \sqrt{2 \over {3 {f_s}}} \left({f \over {f_s}}\right)
^{-7/6} \exp \left[i \Psi_{(I)}(f;f_s,t_c,\xi)\right] 
\ee
for positive frequencies. Above, 
the phase of the 2.5 restricted PN waveform at the $I$-th detector is the 
scalar
\be \label{psiI} 
\Psi_{(I)}(f;f_s,t_c,M,\eta,n_3,n_1) = \vt^{\mu} \xi_{(I)\mu} (f;f_s) \, ,
\ee
with the parameters $\vt^{\mu}$ consisting of the final-coalescence time $t_c$,
the total mass $M$, the mass ratio $\eta (:=m_1 m_2/M^2)$ and the 
source-direction described by two components $n_1$ and $n_3$ of 
unit vector ${\hat n}$ pointing to the source. Given below are 
$\vt^{\mu}$ and $\xi_{(I)\mu}$ :
%angles $(\t,\phi)$, which are 
%expressed in terms of components of unit vector
%$\hat {\sf n}$ pointing towards the source, as given below :
%$\vt^{0} = 2 \pi f_s t_c$, $\vt^{1} = {3 \over {128 \eta}} (\pi M f_s)^{-5/3}$, $\vt^2 = {1 \over {128 \eta}} \left({3715 \over 252} + {55 \over 3} \eta \right) (\pi M f_s)^{-1}$, $ \vt^3 = {{3 \pi} \over {8 \eta}} (\pi M f)^{-2/3}$, $\vt^4 = - {3 \over {128 \eta}} \left( {15293365 \over 508032} + {27145 \over 504} \eta + {3085 \over 72} \eta^2 \right) (\pi M f_s)^{-1/3}$, $\vt^5 = {1 \over {128 \eta}} \left({38645 \over 252} + 5 \eta \right) \pi \ln (f_s)$, $\vt^6 = 2 \pi n_3$, $\vt^7 = 2 \pi n_1$ and
%the frequency dependent functions $\xi_{(I) \mu}$ are
%$\xi_{(I) 0} = \l( {f \over f_s} \r)$, $\xi_{(I) 1} = \l( {f \over f_s} \r)^{-5/3}$, $\xi_{(I) 2} = \l( {f \over f_s} \r)^{-1}$, $\xi_{(I) 3} = \l( {f \over f_s} \r)^{-2/3}$, $\xi_{(I) 4} = \l( {f \over f_s} \r)^{-1/3}$, $\xi_{(I) 5} = \ln \l[ {f \over f_s} \r]$, $\xi_{(I) 6} = z_{(I)} \l({f \over f_s}\r)$ and
%$\xi_{(I) 7} = x_{(I)} \l({f \over f_s}\r)$
{\small{
\bea
\hspace{-1.25cm} \vt^{0} = 2 \pi f_s t_c; && 
\xi_{(I) 0} = \l( {f \over f_s} \r)  \no \\
\hspace{-1.25cm} \vt^{1} = {3 \over {128 \eta}} (\pi M f_s)^{-5/3}; && 
\xi_{(I) 1} = \l( {f \over f_s} \r)^{-5/3} \no \\
\hspace{-1.25cm}\vt^2 = {1 \over {128 \eta}} \left({3715 \over 252} + 
{55 \over 3} \eta \right) (\pi M f_s)^{-1}; && 
\xi_{(I) 2} = \l( {f \over f_s} \r)^{-1} \no \\
\hspace{-1.25cm}\vt^3 = - {{3 \pi} \over {8 \eta}} (\pi M f_s)^{-2/3}; && 
\xi_{(I) 3} = \l( {f \over f_s} \r)^{-2/3} \no \\
\hspace{-1.25cm} \vt^4 = {3 \over {128 \eta}} 
\left( {15293365 \over 508032} + {27145 \over 504} \eta + {3085 \over 72} \eta^2 \right) (\pi M f_s)^{-1/3}; && 
\xi_{(I) 4} = \l( {f \over f_s} \r)^{-1/3} \no \\
\hspace{-1.25cm} \vt^5 = {1 \over {128 \eta}} \left({38645 \over 252} + 
5 \eta \right) \pi ; &~~& 
\xi_{(I) 5} = \ln \l[ {f \over f_s} \r] \no \\
\hspace{-1.25cm} \vt^6 = 2 \pi n_3; && 
\xi_{(I) 6} = z_{(I)} \l({f \over f_s} \r) \no \\
\hspace{-1.25cm} \vt^7 = 2 \pi n_1; && 
\xi_{(I) 7} = x_{(I)} \l({f \over f_s} \r)
\eea
}}
Here, $ x_{(I)}$ and $z_{(I)}$ are, respectively, the $x$ and $z$ coordinate 
values (in units of $c/f_s$) of the location of the $I$-th detector 
in a fiducial reference frame. 

\section{Number of templates}

In this section, we estimate the number of templates required to search over the parameter space. In Ref. \cite{BPD}, we showed 
that for a given pair of source-direction angles $(\t,\phi)$, 
the network likelihood ratio, when maximized over the 
overall amplitude, $\d_c$, $\e$, and $\psi$, gives the network detection
statistic. 
%This statistic turns out to be 
%the norm of network correlation vector projected 
%on the ``helicity'' plane, and is  
%denoted by $\parallel \sf {C}_{\cal H} \parallel$. 
Numerical maximization of the statistic over the rest of the 
parameters, namely, masses and
source-direction angles is performed by using a template bank.
We estimate the number of templates by calculating
the volume of the parameter space of interest obtained by the computing
metric on the manifold and dividing by the size of each template.
When the network statistic is dependent on the parameters solely 
through the difference between the parameter values of the
signal and the template, then 
the metric on the parameter manifold is flat and, hence,
the template placement is uniform. 

It is well known that with PN order $>$ 1,
the metric on the manifold is not flat. The Tanaka-Tagoshi \cite{TAN01}
coordinates provide a convenient and an elegant way to carry out further 
analysis. The salient feature of these coordinates is to make the metric
Euclidean on a flat manifold, which is an approximation to the actual manifold.
Also the coordinate volume of the parameter space in these coordinates is
same as the proper volume which immediately gives the number of
templates as has been described above.

\section{Computational costs}

In this section, we estimate the cost involved in numerically searching over 
the rest of the parameter space mentioned above. This cost has two
important components: 
\bn
\item {{\it The cost involved in FTs} : MLD technique requires
to cross-correlate the data with all possible templates in the rest
of the parameter space involving mass parameters and the direction angles.
Since information of the direction angles is encoded in time delays,
network correlation vectors for templates differing in direction
angles can be constructed by combining the correlation outputs from
different detectors with appropriate time-delays as described in \cite{PDB01}.
Thus, the cost involved in FTs is equal to the number of computational 
operations required in searching over the intrinsic parameters, in our case,
the two masses of the binary.}  
\item {{\it The cost involved in scanning the time-delay window:}
The optimal statistic needs to be evaluated by combining the correlation 
vectors with appropriate time-delays.}
\en

Consider a network of $N_D$ detectors. Let $N$ be the number of sampled points in a data train at each constituent detector. 
If the templates are stored
in memory, then the computing cost in FTs is $6 N_D N n_{X_1-X_2} \log_2 N$, where
$n_{X_1-X_2}$ reflects the number of templates in mass parameters. The number
of floating point operations to construct a network statistic for one pair
of direction angles is $8N_D$. If $n_{tot}$ is the total number of templates
then the total computational cost is
\be
C = 2 N_D N(8 n_{tot} + 3 n_{X_1-X_2} \log_2 N) \, .
\ee
Online data processing requires the data processing rate should be equal to
the data acquisition rate. Thus the length of the data which is effectively processed is equal to the length of the zero padding. We obtain the online computational
speed by dividing the cost by the length of the padding interval. We
use the analytical fits to the noise curves of LIGO and VIRGO which we 
enlist in Table \ref{table:noise}.
\begin{table}[!htb]
\caption{Analytical fits (for positive frequencies) to 
noise power spectral
densities, $2 s_h(f)$, of the interferometric detectors studied in this
paper \cite{SATHYA}. We take $s_h(f)$ to be infinite below the seismic cut-off frequency $f_s$. We 
choose the high frequency cut-off, $f_{c(I)}$, to be $800$ Hz for all $I$.}
\vskip 5pt
{\small{
\begin{tabular}{lcrr}
\hline
Detector & Fit to noise PSD, $10^{46} \times s_h(f) /{\rm Hz} ^{-1}$ &
$f_0$ (Hz) & $f_s$ (Hz)\\
\hline
VIRGO & $ 3.24 \left[(6.23 f/f_0)^{-5}+2(f_0/f)+1+(f/f_0)^2\right]$ & 500 & 20 \\
LIGO I & $9.0\,\left[(4.49 f/f_0)^{-56}+0.16 (f/f_0)^{-4.52} + 0.52 + 0.32 (f/f_0)^2 \right]$ & 150 & 30 \\
\hline
\end{tabular}}}
\label{table:noise}
\end{table}
We tabulate the results for various idealistic as well as realistic networks
in Table \ref{table:cost}. 
\begin{table}[!hbt]
\caption
{Number of templates, computational costs,
and online computing speeds required for a search using specific
networks.  The detector networks are labeled as $I$ for a
single detector, $III$ for three identical detectors with
identical orientations placed on Earth's equator forming an
equilateral triangle.  The detector $X_D$ denotes a
detector with LIGO-I noise at the location of the detector
$D$. The letters $L$, $H$, $V$, $T$, and
$A$ denote, LIGO detector at Louisiana,
LIGO detector at Hanford (of 4 km arm-length),
VIRGO, TAMA and AIGO sites, respectively. We assume LIGO-I noise for
both the LIGO detectors.  We present results for lower mass limits
of $0.5 M_{\odot}$ and $1.0 M_{\odot}$. The maximum length of the $2.5$
PN chirp is $96.5$ secs and $306$ secs for minimal mass limits of  
 $1 M_{\odot}$ and $0.5 M_{\odot}$ respectively.
%The length of
%the restricted $2.5$ PN-chirp is $306$ sec. for $0.5 M_{\odot}$ and 
%$96.5$ sec. for $1 M_{\odot}$. 
We assume fiducial frequency $f_s = 30$ Hz
except for the $LV$ case. We consider data trains of $1100$
sec. for $0.5 M_{\odot}$ and $400$ sec. for
$1.0 M_{\odot}$ sampled at $2$ kHz so that
$N \sim 10^6$.  For the $LV$ network, the length of the
longest chirp is $\sim 284$ sec. for $1.0
M_{\odot}$ and $\sim 900$ sec. for $0.5
M_{\odot}$. The number of points in the data
train $\sim 10^6 - 10^7$. The mismatch is taken to be $3\%$.}
\begin{center}
{\small{
\begin{tabular}{cccccc}
\hline
Network & mass limit& $n_{tot}$ & $n_{X_1-X_2}$ & 
$C_{tot} (\times 10^{14}) $ & $S$ \\
configuration & $(M_{\odot})$& ($\times 10^7)$ 
& ($\times 10^4$) & (fl-pt ops) & (Gflops) \\ 
\hline
$I$      &    $0.5$   &   $0.0214$ & 
$10.7$   &  $0.3$   &${0.37}$\\
         &    $1.0$  &    $0.0042$ &
$2.1$    &  $0.02$ &$0.06$\\ 
$LH$     &    $0.5$  &    $0.81$   &
$16$     & $37.5$  &${4.7}$\\
         &    $1.0$  &     $0.16$  &
$3.2$    & $0.37$  &$1.2$\\ 
$LX_V$   &    $0.5$  &     $2.2$   &
$16$     & $8.7$   &${19}$\\ 
         &    $1.0$  &     $0.44$  &
$3.2$    & $0.72$  &$2.35$\\  
$LX_T$   &    $0.5$  &     $2.67$  &
$16$     & $10$    &${12}$\\ 
         &    $1.0$  &     $0.52$  &
$3.2$    & $0.82$  &$2.7$\\ 
$LX_A$   &    $0.5$  &     $1.7$   &
$16$     & $6.8$  &${15}$\\
         &    $1.0$  &     $0.68$  &
$3.2$    & $1$    &$3.36$\\
$LV$     &   $0.5$   &     $14.1$  &
$93$     & $160$   &${67}$\\
         &   $1.0$   &     $2.7$   &
$19$     & $11$    &$12$\\
$III$    &   $0.5$   &   $1.5 \times 10^3$ &16 & $7.7 \times 10^{3}$&
${9700}$\\
         &  $1.0$   &   $2.9 \times 10^2$  &
$3.2$    & $5.5 \times 10^{2}$&
$180$ \\
\hline
\end{tabular}
}}
\label{table:cost}
\end{center}
\end{table}
For the case of real network of LIGO-VIRGO \cite{BRUCE} with their 
respective noises, we estimate the average number of templates for most 
of the astrophysical
range of $\e$ and $\psi$ to be $n_{tot} \sim {\rm {few~times}} \times 10^{10}$
for lower mass limit of $0.5 M_{\odot}$. We take data trains of length 
$3000$ sec. corresponding to the longest chirp of $\sim 900$ sec. for VIRGO.
Taking a sampling rate of $2$ kHz., the data must be
processed in $2100$ sec. The online data processing demands a 
computational speed of few thousands Tflops.

Here, we note that for networks with $N_D \ge 3$, the computational cost
required to construct optimal network statistic while searching over the
source-direction angles overshoots the FT costs. As a result, the
computational requirements are beyond the reach of the current technology
for a flat search.

\section{Coincident Search}

To focus on the essential aspects of a coincidence search strategy, we 
consider the simplistic case of a network comprising of detectors with
identical noises and orientations, but with arbitrary locations.
The network detection statistic in such a search is taken to be the minimum 
element in $\{|{\cal C}^1|,...,|{\cal C}^{N_D}|\}$, 
where ${\cal C}^I$ is the single detector statistic 
evaluated on the data of the $I$-th detector (see Ref. \cite{Finn}). 
Therefore,
unlike a coherent search, a coincident search involves first establishing 
threshold-crossing by the single-detector detection statistic in each of the 
detectors in a network. Furthermore, claiming a detection by the
network requires that the parameters corresponding to the threshold-crossing 
templates lie within error intervals of one another, such that they can be 
consistently ascribed to a a single astrophysical event. This requirement 
alone immediately implies that, even in this simplistic network the 
computational cost in a coincident search is larger than $N_D C_1$, 
where $C_1$ 
is the computational cost for a single detector search. 

To ascertain exactly how large this cost is, we first 
describe for a network of two identical detectors our search algorithm, which 
is based upon a most powerful search and is not necessarily the cheapest 
computationally \cite{FinnSimilar}:
%First, list the threshold crossers in 
\begin{enumerate}
\item Filter the data $x^1(t)$ and $x^2(t)$ from the two detectors, 
respectively, with a bank of single-detector templates to draw two separate 
%candidate-event lists
lists of threshold crossers. 
Label these ``candidate events'' 
%in detectors 1 and 2 as 
$E^1_i = E(t^{a1}_i;
\mbox{\boldmath $\vartheta$}^1_i)$ and $E^2_j =E(t^{a2}_j;
\mbox{\boldmath $\vartheta$}^2_j)$, 
respectively, where $t^{a1}_{i}$ ($t^{a2}_{i}$) 
denotes the time of arrival of event $i$ at 
detector 1 (2), and
$i=1, 2,...,m$, $j=1, 2,...,n$. Note that $m \neq n$, in general.
Also, $\mbox{\boldmath $\vartheta$}^1_i$ denotes the template-parameter vector 
characterizing event $i$ at detector 1. The above nomenclature is suited to
handle the 
possibility of
two or more templates triggering off simultaneously, say, on the data from 
detector 1. In such a case, one will have more than one event 
with $t^{a1}_{i-1}=t^{a1}_i=t^{a1}_{i+1}$, but with 
$\mbox{\boldmath $\vartheta$}^1_{i-1} \neq
\mbox{\boldmath $\vartheta$}^1_i \neq \mbox{\boldmath $\vartheta$}^1_{i+1}$. 

\item {``Time window'' veto:}   
%{\bf Execute coincidence-window vetoing:} 
Let detector 1 have a smaller number of candidate events than detector 2.
%labeled $E^1_i$. With each $E^1_i$, in detector 1, 
With each $E^1_i$ associate a set $W^2(t^{a1}_i; \tau^{12}_{c~ij})$ of 
candidate events $E^2_j$, such that 
$t^a_i - \tau^{12}_{c~ij} \leq t^a_j \leq t^a_i + \tau^{12}_{c~ij}$. 
Here, $\tau^{12}_{c~ij}$ is the sum of the light-travel time between the two 
detectors and the sum of magnitudes of 
the estimated errors in their arrival times at detectors 1 and 2. 
Note that an event $E^2_j$ may appear in more than one set. That is, it
may happen that $E^2_j \in W^2(t^{a1}_i; \tau^{12}_{c~ij}) 
\cap W^2(t^{a1}_k; \tau^{12}_{c~kj})$, where $i\neq k$. Discard from the lists
those $E^2_j$ that
do not belong to any $W^2(t^{a1}_i; \tau^{12}_{c~ij})$. 

\item {``Parameter window'' veto:} 
Compute the covariance matrix in the parameter space around $E^1_i$ and 
around each event in $W^2(t^{a1}_i; \tau^{12}_{c~ij})$ from the ambiguity 
function \cite{Hels}. Estimate the parameter error, 
$\Delta\mbox{\boldmath $\vartheta$}^1_i$ 
($\Delta\mbox{\boldmath $\vartheta$}^2_j$), to be the square-root of the 
variance of the parameter $\mbox{\boldmath $\vartheta$}^1_i$ 
($\mbox{\boldmath $\vartheta$}^2_j$) derived from this matrix. 
Discard events in $W^2(t^{a1}_i; \tau^{12}_{c~ij})$ that
have  $|\vartheta^{2~\mu}_j-\vartheta^{1~\mu}_i| > 
|\Delta\vartheta^{2~\mu}_j| + |\Delta\vartheta^{1~\mu}_i|$ for each parameter
index $\mu$.
\end{enumerate}
\noindent The pairs of candidate events surviving the above vetoes are the 
``detected'' events. A more sophisticated approach involving further vetoes of 
the type discussed in Ref. \cite{AllenEtAl} will be studied elsewhere. 

The above steps explicitize the computational costs, over and above that of
$N_D C_1$, that are necessary in a coincidence detection, but are often 
glossed over: {\em Extra costs are involved in computing parameter errors and implementing vetoes based on them.}
These costs obviously scale as the 
number of the candidate events in each detector (whereas, the cost in
a coherent search is independent of it). These counts, in
turn, depend on the value of the detection threshold and, therefore, on the
false-alarm probability. The number of floating point operations (Flop)
needed to estimate the error in a parameter, $\vartheta^{I~\mu}_i$, is close to that involved in 
taking the second derivative of ${\cal C}^I$ with respect to 
$\vartheta^{I~\mu}_i$. 
Using the discrete version of second derivative the number of Flop involved
$\sim 10C_1$. Therefore, in an $8$ dimensional parameter space 
(based on the independent parameters 
$(r, \delta_c, \vartheta^0,...,\vartheta^5)$, the 
number of Flop required to estimate parameter errors for all candidate 
events is about $80 C_1\times\sum_{I=1}^{N_D} N_I$, where $N_I$ is the
number of candidate events in detector $I$. Additional operations
required to compare the parameter values across detectors (using the inequality
given in step (iii) above) and veto events scales as $\prod_{I=1}^{N_D} N_I$,
which is a small fraction of the total cost for $N_D={\cal O}(1)$ and
$N_I={\cal O}(10)$. Thus, neglecting this last contribution, 
the total number of Flop scales as:
\be\label{coinOps}
N_D C_1 + 80 C_1~\sum_{I=1}^{N_D} N_I \ \ .
\ee
For comparison with the coherent search costs, we take $N_I = 10^2$ in 
$N\simeq 10^6$ data points in each of the 3 detectors in a 
network. For, a minimum
mass of $0.5 M_\odot$, Table 2 shows that $C_1 = 0.3\times 10^{14}$. Thus,
for network configuration $III$, the total number of Flop in a 
coincident search is about 7.2$\times 10^{17}$ which is very close to 
$C_{tot} = 7.7\times 10^{17}$ for a coherent search. 
One may argue that it is possible to reduce $N_I$
in each detector by using additional vetoes of the type adopted in Ref.
\cite{AllenEtAl}. Such steps will surely reduce the contribution 
from the second term in eq.(\ref{coinOps}).
Nevertheless, the additional costs in implementing such
vetoes is very large as well and must be explored in more detail.

It is easy to see that with more events or more 
detectors, the cost related to Eq. (\ref{coinOps}) can only rise.
With future detectors, where the detection thresholds
will be lower than present ones owing to their higher sensitivity, the number 
of events with larger signal-to-noise ratios will increase, consequently,
increasing the computational cost further of a coincidence search.
 
\section{Conclusion}
As shown in Table 2., the computational cost in a coherent search rises
markedly in going from $N_D=1$ to 3. This is expected because the number of
parameters and, therefore, the parameter volume accessible to a search 
increases from 5 (for $N_D=1$) to 9 (for $N_D=3$). Indeed, for $N_D \geq 3$ the
cost required to search over source-direction angles overshoots
that required for the FFTs. Beyond $N_D=3$, the computational cost in a 
coherent search, however, stabilizes. This must be contrasted
with the cost behavior in a coincidence search, where it continues to 
increase with $N_D$. Specifically, given a network of identical detectors and 
a false-alarm probability, for a low enough detection threshold, a coincidence 
search will cost more than a coherent search for $N_D>1$. In either search,
the computational costs are very large and, hence, call for 
investing in the exploration of more efficient search techniques, such as
hierarchical strategies. 
\ack
%\noindent {\bf Acknowledgement:}
AP would like to thank CSIR, India for SRF.


\bigskip

\begin{thebibliography}{9}
\bibitem{DIB01} 
T.~Damour, B.~R.~Iyer and B.~S.~Sathyaprakash,\/ Phys.\ Rev.\ D\ {\bf 63}, 044023 (2001).

\bibitem{PDB01}
A.~Pai, S.~V.~Dhurandhar and S.~Bose,\/ Phys.\ Rev.\ D\ {\bf 64}, 042004 (2001).

\bibitem{Hels} 
C.~W.~Helstrom, {\sl Statistical Theory of Signal Detection} (Pergamon Press,
London, 1968).

\bibitem{BPD}
S.~Bose, A.~Pai, and S.~V.~Dhurandhar, \/ Int.\ J.\ Mod.\ Phys. \ D\
{\bf 9}, 325 (2000). (gr-qc/0002010)

\bibitem{TAN01}
T.~Tanaka and H.~Tagoshi,\/ Phys.\ Rev.\ D\ {\bf 62}, 0822001 (2000).

\bibitem{BRUCE}
B.~Allen, {\it Gravitational wave detector sites}, gr-qc/9607075.

\bibitem{SATHYA}
Analytical fits obtained by private communication with Dr. B. S. Sathyaprakash.

%\bibitem{JK}
%P.~Jaranowski, A.~Krolak, K.~D.~Kokkotas, and G.~Tsegas,\/ Class.\ Quant.\ 
%Grav.\ {\bf 13}, 1279 (1996).

\bibitem{Finn}
S.~Finn,\/ Phys.\ Rev.\ D{\bf 63}, 102001 (2001). (gr-qc/0010033)

\bibitem{FinnSimilar}
This algorithm extends the one described in Ref. \cite{Finn}.

\bibitem{AllenEtAl} 
B.~Allen {\it et al}, Phys.~Rev.~Lett.~{\bf 83} 1498 (1999). (gr-qc/9903108)

\end{thebibliography}
\end{document}